\documentclass[10pt,aps,prl,superscriptaddress,
twocolumn,showpacs,floatfix,notitlepage]{revtex4-1}
\usepackage{amsmath}
\usepackage{amsfonts}
\usepackage{amssymb}
\usepackage{hyperref}
\usepackage[shortcuts]{extdash}
\usepackage{graphicx}
\usepackage{color}
\usepackage{xcolor}
\usepackage[mathscr]{euscript}
\usepackage{natbib}
\usepackage{rotating}

\hypersetup{colorlinks=true,linkcolor=magenta
  ,citecolor=magenta, urlcolor=magenta}

\begin{document} 
\flushbottom 

\title{Exceptional points and phase transitions in non-Hermitian binary systems}

\author{Amir Rahmani}
\affiliation{%
	Institute of Physics Polish Academy of Sciences, Al. Lotnik\'{o}w 32/46, 02-668 Warsaw, Poland}%
 \author{ Andrzej Opala}
\affiliation{%
	Institute of Physics Polish Academy of Sciences, Al. Lotnik\'{o}w 32/46, 02-668 Warsaw, Poland} 
 \affiliation{%
 Institute of Experimental Physics, Faculty of Physics, University of Warsaw, ul. Pasteura 5, PL-02-093 Warsaw, Poland}
\author{ Micha\l{} Matuszewski}
\affiliation{%
	Institute of Physics Polish Academy of Sciences, Al. Lotnik\'{o}w 32/46, 02-668 Warsaw, Poland}%

\date{\today}

\begin{abstract}
 Recent study demonstrated that steady states of a polariton system may demonstrate a first-order dissipative phase transition with an exceptional point that appears as an endpoint of the phase boundary~[R. Hanai et al., Phys. Rev. Lett. 122, 185301 (2019)]. Here, we show that this phase transition is strictly related to the stability of solutions. In general, the exceptional point does not correspond to the endpoint of a phase transition, but rather it is the point where stable and unstable solutions coalesce. Moreover, we show that the transition may occur also in the weak coupling regime, which was excluded previously. In a certain range of parameters, we demonstrate permanent Rabi-like oscillations between light and matter fields. Our results contribute to the understanding of nonequilibrium light-matter systems, but can be generalized to any two-component oscillatory systems with gain and loss.  
\end{abstract}

\maketitle

Phase transitions correspond to significant alterations of the properties of a system caused by the modification of physical parameters. 
Examples include the ferromagnetic-paramagnetic phase transition~\cite{DavidJ15}, gas-liquid-solid transition~\cite{sergey18}, Bose-Einstein condensation in bosonic and fermionic systems~\cite{proukakis2017universal}, metal–insulator transition in solid state~\cite{RevModPhys.70.1039}, and topological phase transitions~\cite{brink2018topological}. Phase transitions may also occur in non-Hermitian systems,  which are systems that do not satisfy the condition of Hermiticity, which is embedded in quantum mechanics~\cite{Ashida20}. Here the non-Hermitian contributions may stem from dissipation~\cite{Miri19} or asymmetric coupling~\cite{PhysRevX.8.031079} and lead to a number of unique properties such as non-reciprocity~\cite{pp14},  mutually interlinked non-Hermitian phase transitions~\cite{Weidemann22} and the non-Hermitian skin effect~\cite{okuma23}.

    A striking example of non-Hermitian physics that deviates significantly from the Hermitian case is the coalescence of eigenstates and energy eigenvalues at so-called exceptional points (EPs). These spectral singularities may be accompanied by a non-Hermitian phase transition~\cite{Fahri21}. Standard procedure to investigate these phase transitions is through the study of the spectrum of the system as some controllable parameters  are changed~\cite{Miri19}. Typically, the process involves meticulous adjustment of loss and gain in order to achieve the desired outcome. In general, in a linear system the presence of EPs is independent of the stability of the stationary state that the system evolves to~\cite{Khurgin:20}.  However, in a nonlinear system, more than one solution may be stable, which gives rise to the phenomena of bistability and multistability~\cite{boyd2020nonlinear,paraiso2010multistability,gippius2007polarization,cancellieri2011multistability}. The existence of nonlinear features may affect the non-Hermitian effects realized in linear cases or give rise to entirely new phenomena~\cite{Yu21,jan23,PhysRevA.103.043510,xia2021nonlinear,Absar2015,Ramezani2010,el2018non,wimmer2015observation}.
    
In order to examine the relationship between nonlinearity and non-Hermitian physics, it is necessary to study systems that possess variable nonlinearity and controllable gain and loss. 
Particularly suitable systems for this study are those where matter couples with light, as they allow to take advantage of the difference in physical properties of these components. For example, it was demonstrated that exceptional points appear naturally in light-matter systems of exciton-polaritons and subtreshold Fabry-Perot lasers~\cite{Khurgin:20,Hanai19a}. Moreover, it is possible to induce exceptional points by manipulating spatial and spin degrees of freedom of exciton-polaritons in various configurations~\cite{jan23,gao2018chiral,Gao2022,krol2022annihilation,gao2015observation,su2021direct,song2021room,hanai2020critical,rahmani2023non,opala2023natural,gao2018continuous,liao2021experimental}. In the case of bosonic condensates of exciton-polaritons, it was predicted that a dissipative first-order phase transition line exists in the phase diagram~\cite{Hanai19a}, similar to a critical point in a liquid-gas phase transition. According to this study, this phase transition line exists in the regime of strong light-matter coupling and has an endpoint which corresponds to an exceptional point~\cite{Hanai19a}. 


In this letter, we investigate a non-Hermitian model describing interaction between two oscillating modes. We use it to examine the significance of nonlinearity in a non-Hermitian phase transition. This model can describe light and matter modes in exciton-polariton condensation and lasing, as investigated in Ref.~\cite{Hanai19a}. We find that the model is incomplete unless nonlinear saturation of gain is taken into account. Importantly, saturation increases the complexity of the phase diagram and leads to the appearance of  bistability. It has also profound consequences on the physics of the system. We find that while the first-order phase transition line with an endpoint is present, the equivalence of the endpoint to an exceptional point as found in~\cite{Hanai19a} is no longer valid in the general case. The phase diagram of Ref.~\cite{Hanai19a} can be restored in the limit of strong saturation. In contrast to the results of Ref.~\cite{Hanai19a}, the transition between solutions can occur also in the weak coupling regime. This suggests that the second threshold from polariton to photon lasing, observed in experiments~\cite{Hu21,Pieczarka22,Tempel12}, may be related to a dissipative phase transition in the weak coupling regime. Moreover, we find a regime of permanent Rabi-like oscillations between two stable solutions. This regime corresponds to a line in the phase diagram that ends with an exceptional point.

%
%
\emph{Model and Analytical Solutions.} We consider a system of two coupled oscillators described by a non-Hermitian Hamiltonian with gain and loss. The imbalance between gain and loss in a linear system leads in general to solutions exponentially growing or decaying in time. To obtain non-trivial stationary solutions it is necessary to include nonlinearity. Here we adopt cubic nonlinearity that appears naturally in symmetric systems with no dependence on the complex phase. Such a model can be realized, among many other physical systems, in the case of cavity photons coupled to excitons, where the nonlinearity occurs only in the matter (exciton) component~\cite{Kavikin16}. The system is described by complex functions $\psi_C=n_Ce^{i\varphi_C}$ and $\psi_X=n_Xe^{i\varphi_X}$, corresponding  to amplitudes of cavity photons and excitons, respectively.
The dynamics is governed by equations 
$i\hbar\partial\psi/\partial t = $ 
$i\hbar \partial_t|\Psi\rangle=H|\Psi\rangle$ with  $|\Psi\rangle=(\psi_C,\psi_X)^\mathrm{T}$, where non-Hermitian Hamiltonian $H$ is given by~\cite{Hanai19a}
\begin{align}\label{eq:ejhr984yrhrf8erfer8f}
H=\left( \begin{matrix}
E_C-i\hbar\gamma_C & \hbar\Omega_R\\
\hbar\Omega_R & E_X+g|\psi_X|^2+ip
\end{matrix}\right)\,.
\end{align}
Here $\hbar\Omega_R$ is the coupling strength, $\gamma_C$ is the decay rate of the photon field, and $p$ represents the gain to the exciton field. This gain can be realized in practice by nonresonant optical or electrical pumping. We define the complex nonlinear coefficient as $g=g_1-ig_2$, where $g_1$ is the strength of two body interactions (Kerr-like nonlinearity) and $g_2|\psi_X|^2$ is the saturation term that allows to avoid instability. Spectrum of Hamiltonian~(\ref{eq:ejhr984yrhrf8erfer8f}) can be found analytically
\begin{align}\label{eq:djhf7834yr2df}
E=&\frac12 \big[E_c+\mathcal{E}+i(\mathcal{P}-\hbar \gamma_c)\nonumber\\& \pm \sqrt{4\hbar^2\Omega_R^2+[\mathcal{E}-E_c+i(\mathcal{P}+\hbar\gamma_c)]^2}\big]\,,
\end{align}
where $\mathcal{P}=p-g_2(n_X^{\mathrm{SS}})^2$ and $\mathcal{E}=E_x+g_1 (n_X^{\mathrm{SS}})^2$. For  convenience, we denote the solution associated with plus (minus) by $U(L)$. The respective steady state analytical solutions $|\Psi\rangle=|\Psi_0\rangle e^{-i E t}$ 
can be found from the condition $\mathrm{Im}[E]=0$, that is, the imaginary part of the eigenvalue of~(\ref{eq:ejhr984yrhrf8erfer8f}) must be zero. In~\cite{Hanai19a}, it was argued that   one or two real energy solutions exist in certain regions in parameter space. However, it can be seen from (\ref{eq:djhf7834yr2df}) that except from special values of parameters, real energy solutions can exist only when saturation represented by $g_2$ is taken into account. 
We will show below that accounting for the nonlinear $g_2$ term does in fact lead to the appearance of up to three real-energy solutions, each of them of the form~(\ref{eq:djhf7834yr2df}).

The condition $\mathrm{Im}[E]=0$ allows one to find analytical expression for $n_X^{\text{SS}}$
\begin{align}
(n_X^{\mathrm{SS}})^2=\frac{1}{g}\big(\mathrm{Re}[E]-E_X-iP-\frac{(\hbar\Omega_R)^2}{\mathrm{Re}[E]-E_C+i\hbar\gamma_C}\big).
\end{align}
The resulting explicit formula for $n_X^\text{SS}$ is tedious, but for a given $n_X^\text{SS}$, one can find closed forms of steady state  $n_C^\text{SS}$ and $\varphi_{CX}=\varphi_C-\varphi_X$
\begin{align}
n^{\mathrm{SS}}_C=&n^{\mathrm{SS}}_X\sqrt{\frac{p}{\hbar\gamma_C}-\frac{\left(n_X^{\mathrm{SS}}\right)^2g_2}{\hbar\gamma_C}}\,,\\
\varphi_{CX}^{SS}=&\arg\left(\frac{\delta-g_1(n_X^{\mathrm{SS}})^2}{\hbar\Omega_R\left(n^{\mathrm{SS}}_C/n_X^{\mathrm{SS}}-n_X^{\mathrm{SS}}/n^{\mathrm{SS}}_C\right)}-i\frac{\gamma_C n^{\mathrm{SS}}_C}{\Omega_R n_X^{\mathrm{SS}}}\right)\,,
\end{align}
where we introduced photon-exciton energy detuning $\delta=E_C-E_X$.

\begin{figure}
 	\begin{center}
 	\includegraphics[width=\columnwidth]{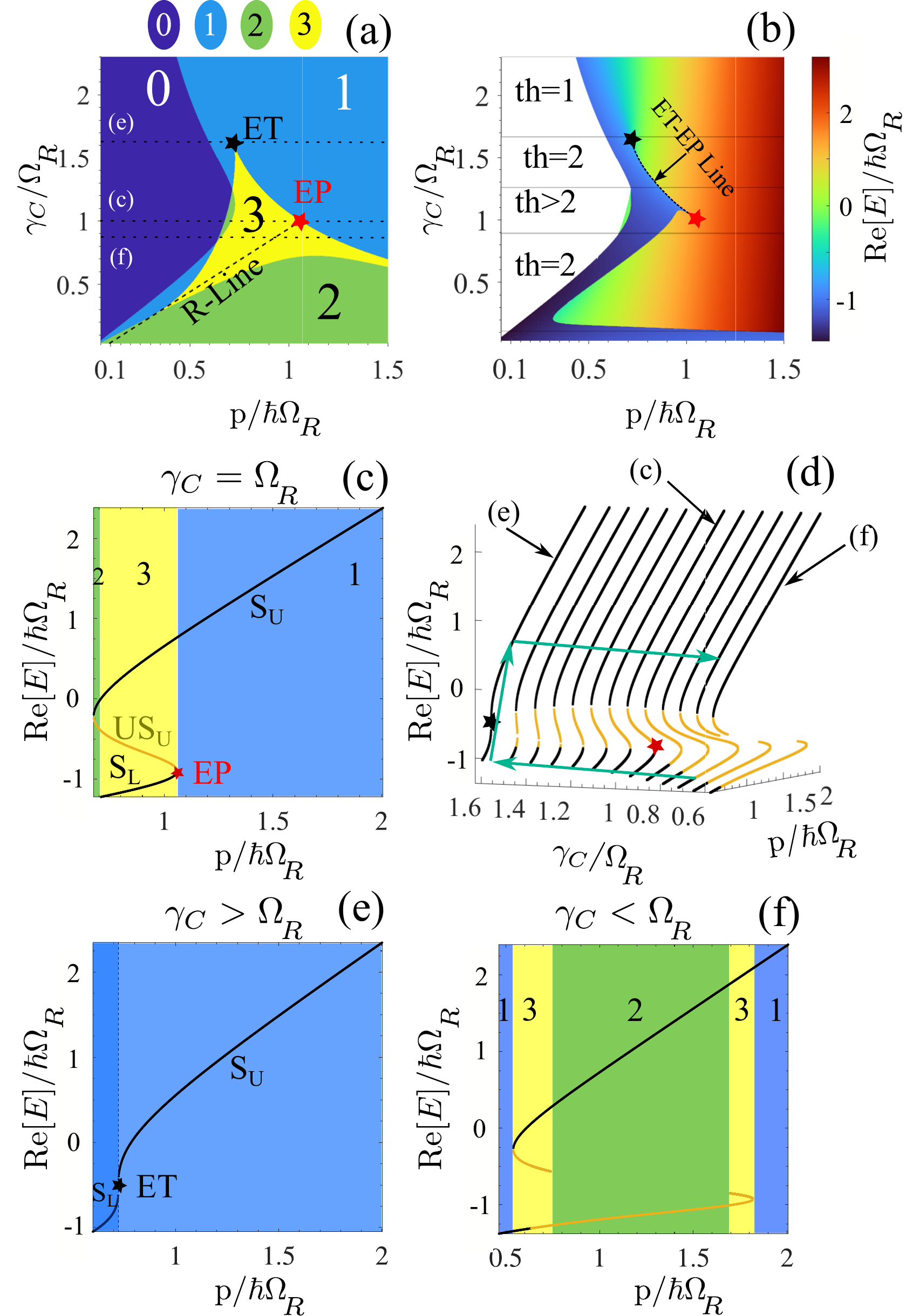}
 	\caption[]{ (a-b): Phase diagrams of generic nonresonantly driven binary system~(\ref{eq:ejhr984yrhrf8erfer8f}). In (a) the number of stationary states is marked with colors in the function of normalized photon decay rate ($\gamma_c$) and pumping strength ($p$). In (b) only the lowest-energy stable state is shown. Here colors indicate the real part of  the energy of the corresponding solution. In (a) and (b) the exceptional point (EP, marked with a red star) appears both as a point on the phase boundary and as an endpoint of the R-line. The R-Line corresponds to non-decaying Rabi-like oscillations between two modes. The endpoint of the phase transition (ET) is marked with a black star. Cross-sections of constant $\gamma_c$ with different numbers of thresholds (th) are marked with horizontal lines. Panels (c-f) show bistability and phase transitions in more detail. In (c) we show the case $\gamma_C=\Omega_R$, for which the energy eigenvalues coalesce at the EP, which is also a turning point of a bistability curve. 
  Stable solutions are marked with S and black lines, while unstable solutions are marked with US and orange lines.
  Panel (d) shows real part of energies for different pumping and decay rates. The ET point corresponds to the transition to bistability at $\gamma_C>\Omega_R$. This cross-section is depicted in panel (e), while in panel (f)  we show the case $\gamma_C<\Omega_R$, where the unstable solution is split into two branches, and the lowest-energy solution becomes unstable in certain regions of pumping. Other parameters are $\delta=0.2~\hbar\Omega_R$, $g_1=0.1~\hbar\Omega_R$, $E_X=0$, $E_C=0.2~\hbar\Omega$ and $g_2=0.3~g_1$.}
 	\label{fig:1}
 	\end{center}
 \end{figure}
\emph{Non-Hermitian Phase Transitions.}
We use the analytical solutions from the previous section to determine the phase diagram of the system, looking at it from two perspectives. We analyze the steady state solutions and their multiplicity, as in Fig.~\ref{fig:1}(a). On the other hand, we consider the lowest-energy state among the dynamically stable ones and investigate its properties and possible transitions, see Fig.~\ref{fig:1}(b). The latter approach is equivalent to analyzing a system that is weakly coupled to an energy sink, which does not perturb the spectrum, but picks the lowest-energy stable solution after a sufficiently long evolution due to its energetic stability.

In the case when the conservative nonlinearity $g_1$ is stronger than the dissipative nonlinearity $g_2$, representative phase diagrams are shown in Fig.~\ref{fig:1}. We focus on the blue-detuned case ($\delta>0$), which is much richer that the red-detuned case. In Fig.~\ref{fig:1}(a) the number of steady state solutions is shown. Up to three non-zero solutions, corresponding to both upper and lower branches of Eq.~(\ref{eq:djhf7834yr2df}) can exist, which results from the nonlinearity of the system. The region of zero solutions corresponds to the situation where pumping cannot overcome losses and no lasing nor polariton condensation occurs. For given $\Omega$ and $\gamma_C$, increasing pumping $p$ can lead to one or several thresholds, as indicated with horizontal lines.

Special points in the phase diagram (marked by stars in Fig.~\ref{fig:1}) include the exceptional point (EP) and the endpoint of the first-order phase transition (ET). In contrast to~\cite{Hanai19a}, we find that in general they do not coincide. To determine the position of the EP, one can find the following conditions for which the real and imaginary parts of eigenvalues are zero in Eq.~(\ref{eq:djhf7834yr2df})
\begin{equation}\label{eq:ouedjdyedhu9qw0eyhd}
p^{\mathrm{EP}}=\hbar\Omega_R+\frac{g_2\delta}{g_1}\,,~~\gamma_C=\Omega_R\,.
\end{equation}
This can occur when $n_X^{\mathrm{SS}}=\delta/g_1$, that is, whenever the system is blue-detuned ($\delta>0$). 
On the other hand, the ET point is clearly visualised in the phase diagram that takes into account the energetic instability in panel Fig.~\ref{fig:1}(b). The first-order phase transition line begins at the ET point in the weak coupling regime ($\gamma_C>\Omega_R$) and follows the arc represented by the ET-EP line towards the EP point. Below the EP, the phase transition line follows into the strong coupling regime. We conclude that, contrary to the results of~\cite{Hanai19a}, the first-order phase transition can occur also in the weak coupling regime. This can be explained by a simple physical argument. Since the pumping influences the effective photon-exciton detuning $\tilde{\delta}=E_C-(E_X+g (n^{\mathrm{SS}}_X)^2)$, the increase of pumping can change of the sign of $\tilde{\delta}$, leading to an abrupt change of the lowest-energy state in the weak-coupling regime.

Figure~\ref{fig:1}(d) shows the dependence of the real part of the energy of solutions shown in Figs.~\ref{fig:1}(a,b), in the vicinity of the ET-EP line. As can be seen, the ET point is the point of the transition to bistability. On the other hand, the EP point corresponds to a turning point in the bistability curve. The cross-section including the EP point ($\gamma_C=\Omega$) is depicted in more detail in Figure~\ref{fig:1}(c), which shows the occurrence of two stable branches from the upper and lower branches of Eq.~(\ref{eq:djhf7834yr2df}) and one unstable branch. At the EP, the unstable upper branch coalesces with the lower stable branch, leading to the first-order phase transition. The cross-section with the ET point ($\gamma_C>\Omega_R$) is shown in Fig.~\ref{fig:1}(e), where the bistability curve closes, and the transition from the upper to lower branch becomes smooth. This leads to the possibility to encircle the exceptional point as indicated with arrows in Fig.~\ref{fig:1}(d). 

Interestingly, additional features that have an influence on the physics of the system can occur in the strong coupling case ($\gamma_C<\Omega_R$), see Fig.~\ref{fig:1}(f). These include the disappearance of one of the solutions in a certain parameter range and the dynamical instability of the lowest-energy branch (marked with orange line). Consequently, the upper, higher-energy solution may become the only viable solution despite the existence of lower-energy solutions.

\begin{figure}
 	\begin{center}
 	\includegraphics[width=\columnwidth]{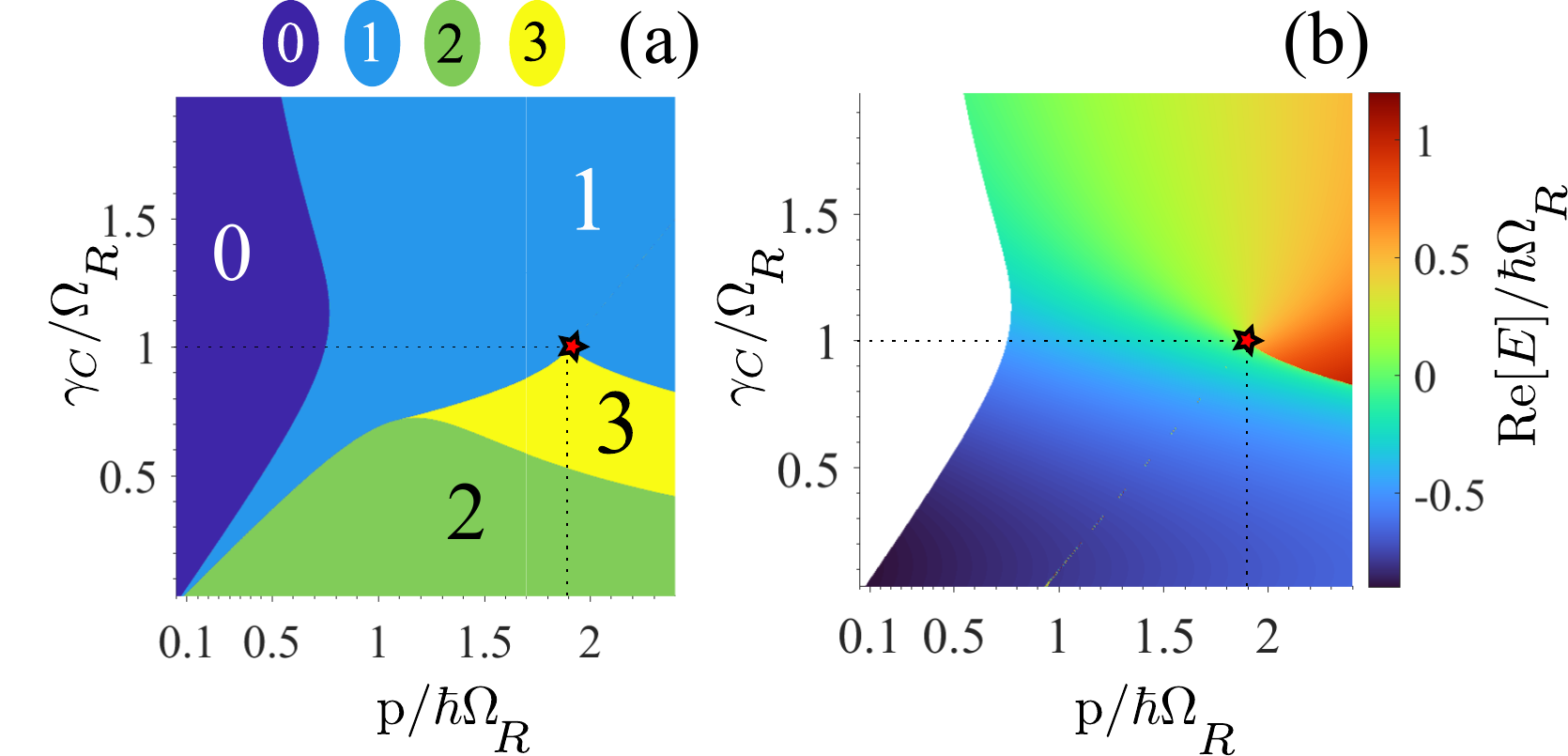}
 	\caption[]{Phase diagrams in the case when dissipative nonlinearity $g_2$ dominates over the conservative nonlinearity $g_1$. In this case, endpoint of the phase transition (ET) and exceptional point (EP) correspond to the same point in parameter space, recovering the results of~\cite{Hanai19a}. Parameters are the same as in Fig.~\ref{fig:1}, except for $g_2=4.5~g_1$.}
 	\label{fig:2}
 	\end{center}
 \end{figure}
In the opposite case when the dissipative nonlinearity dominates over the conservative one, we find that the phase diagram of energetically stable solutions recovers the results of~\cite{Hanai19a}, see Fig.~\ref{fig:2}. As the dissipative nonlinearity is increased, the length of the ET-EP arc decreases, and finally the two points coalesce.  In this specific case, the exceptional point is characterized by a jagged crest in the phase diagram, embodying a third-order exceptional point (see supplementary materials). This phenomenon arises from the coalescence of two stable solutions and a single unstable solution.

\begin{figure}[htb]
 	\begin{center}
 	\includegraphics[width=\columnwidth]{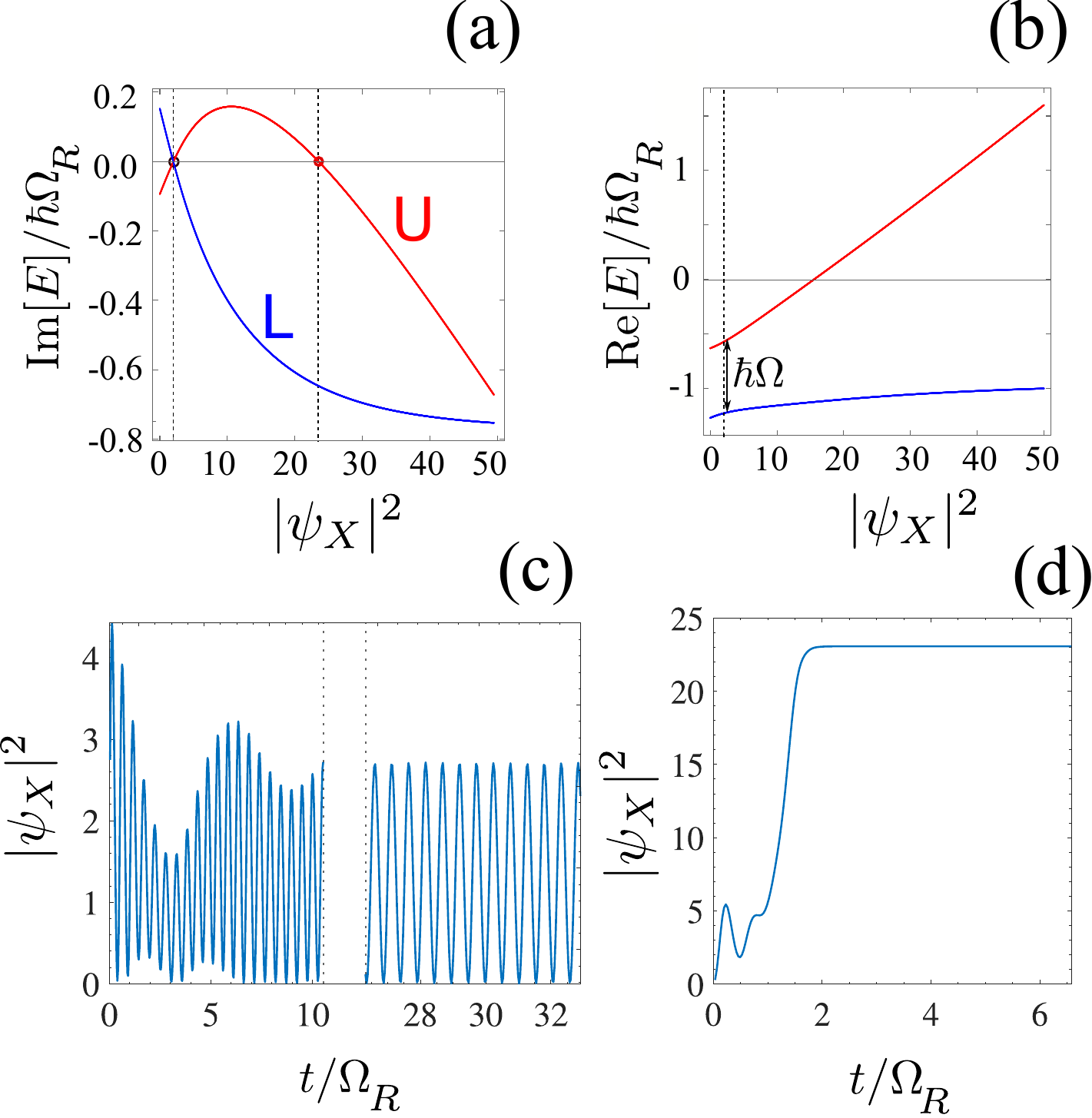}
 	\caption[]{Permanent oscillations. (a) Imaginary part of energy eigenvalues versus exciton density $n_X$. The condition $\mathrm{Im}[E(n_X)]=0$ provides the possible stationary solutions. Here the solution from the lower energy branch (denoted by L) coincides with one of the solutions from the upper branch (denoted by U). (b) Real part of eigenvalues versus $n_X$. The nonzero energy gap between the solutions results in Rabi oscillations with frequency $\Omega_R$. Examples of simulations are shown in (c) and (d). In (c) oscillations stabilize after initial transient. In (d), different initial condition for the same parameters leads to a steady state solution from the upper branch. Using parameters: $p=0.82~\hbar \Omega_R$, $\gamma_C=0.75~\Omega_R$, $g_1=0.1~\hbar\Omega_R$, $g_2=0.3~g_1$ and $\delta=0.2~\hbar\Omega_R$.}
 	\label{fig:3}
 	\end{center}
 \end{figure}

\emph{Permanent Rabi-Like Oscillations: R-Line.}
Our analysis allows to predict that a peculiar oscillating state may form, as indicated in Fig.~\ref{fig:1}(a) by \emph{R-Line}.  In this case, long evolution leads to permanent oscillations, resembling Rabi oscillations in a two-level system, instead of stationary solutions. To explain this phenomenon, we examine imaginary and real parts of eigenvalues given in Eq.~(\ref{eq:djhf7834yr2df}). An example is shown in Figs.~\ref{fig:3}(a) and~\ref{fig:3}(b). 
In general, two kinds of stationary solutions corresponding to $\mathrm{Im}[E(n_X)]=0$ may exist. As shown in  Fig.~\ref{fig:3}(a), in this particular case there are two solutions from the upper branch and one solution from the lower branch (the black dashed vertical lines denote
the emergent solutions). Our interest is in solutions from upper and lower branches that occur at the same $n_X$, while there is a gap in respective real parts, see  Fig.~\ref{fig:3}(b). Such solutions occur when $p=(g_2/g_1)\delta+\hbar\gamma_C$, which corresponds to a straight line (marked by R-line) in the phase diagram of Fig.~\ref{fig:1}(c). 

An example of such permanent oscillations is shown in Fig.~\ref{fig:3}(c). After initial transient time, the oscillations stabilize at a cetain amplitude. When different initial conditions are used, the system may end up in one of the steady state solutions, as shown in Fig.~\ref{fig:3}(d). The frequency of oscillations is given by the gap, $\Omega=2\sqrt{\Omega_R^2-\gamma_C^2}$. When the parameters of the system approach the exceptional point along the R-line, the gap decreases and the period of oscillations increases. At the exceptional point ($\Omega_R=\gamma_C$), the solutions coalesce and the period becomes infinite. Therefore, the exceptional point is the endpoint of the R-line. 
%
%


{\it Discussion}. We showed that, contrary to previous understanding, non-Hermitian polariton systems exhibit first-order phase transition with an endpoint that in general does not coincide with the exceptional point. Explanation of this phenomenon requires taking into account the nonlinear gain saturation and the consideration of the bistability curve. While the endpoint of the phase transition is where the bistability appears, the exceptional point is where the stable and unstable solutions coalesce. In addition, we demonstrated that first-order phase transition may occur in the weak coupling regime, and that for certain values of parameters one can predict permanent oscillations, whose frequency vanishes at the exceptional point.

The predicted results contribute to the ongoing debate surrounding polariton/photon lasing. The presence of an exceptional point has been identified as the possible underlying factor for the observed second threshold ~\cite{Hanai19a}. Here, we provide further insights by identifying several other thresholds in phase diagrams and pointing out that multiplicity and stability of solutions are also crucial factors, so far overlooked. 

The presented results may be applied to much broader class of systems. The non-Hermitian Hamiltonian represented by the $2\times2$ matrix in Eq.~(\ref{eq:ejhr984yrhrf8erfer8f}) describes in general an arbitrary two-mode oscillatory system with gain and loss in the two modes, and the cubic nonlinearity in one of them. This term appears naturally in any oscillatory system in the first order as long as the nonlinearity respects the global $U(1)$ symmetry of the oscillations. Examples include not only all quantum mechanical systems such as Bose-Einstein condensates, but also high-frequency coupled classical oscillators, where phase of oscillations is irrelevant on the time scale of a slowly varying envelope. The results presented here should be applicable to any such system that exhibits exceptional points and nonlinearity.

A.R. and M.M. acknowledge support from National Science Center, Poland (PL), Grant No. 2016/22/E/ST3/00045. A.O. acknowledges support from Grant No. 2019/35/N/ST3/01379.

\bibliography{EP}

\end{document}